# Collective ferromagnetism of artificial square spin ice

N. S. Bingham[1], X. Zhang[1], J. Ramberger[2], O. Heinonen[3], C. Leighton[2], P. Schiffer[1,4]

1. Department of Applied Physics, Yale University, New Haven, CT 06511 USA
2. Department of Chemical Engineering and Materials Science, University of Minnesota, Minneapolis, Minnesota 55455, USA
3. Materials Science Division, Argonne National Laboratory, Argonne, IL, 60439, USA,
4. Department of Physics, Yale University, New Haven, CT 06511 USA

**Abstract:**

We have studied the temperature and magnetic field dependence of the total magnetic moment of large-area permalloy artificial square spin ice arrays. The temperature dependence and hysteresis behavior are consistent with the coherent magnetization reversal expected in the Stoner-Wohlfarth model, with clear deviations due to inter-island interactions at small lattice spacing. Through micromagnetic simulations, we explore this behavior and demonstrate that the deviations result from increasingly complex magnetization reversal at small lattice spacing, induced by inter-island interactions, and depending critically on details of the island shapes. These results establish new means to tune the physical properties of artificial spin ice structures and other interacting nanomagnet systems, such as patterned magnetic media.

Artificial spin ice systems (ASI) [1] consisting of two-dimensional arrays of ferromagnetic single-domain nanoislands can be studied in a nearly limitless range of lattice geometries that lead to exotic collective behavior [2–5]. Control over the design of the lattice geometry has enabled experimental study of a range of physical phenomena, including classical statistical physics models, magnetic-monopole-like excitations, and unusual topological physics [4,5], and possible applications, including novel computing paradigms and magnonic devices [6,7]. ASI studies typically treat the individual ferromagnetic elements as simple Ising-like moments that switch between orientations with thermal fluctuations or upon application of magnetic field. The reversal of island moments is recognized, however, to have considerable complexity [8–15], with dependence on island size and shape as well as the lattice spacing and geometry [9,16–19]. Notably, ASI also serves as an accessible model platform for probing superparamagnetism with an unusually high degree of control over the moments and their interactions [20].

Despite extensive recent attention, only a small number of researchers have examined the collective static magnetization of entire ASI arrays [21–30], due to the large array dimensions required. Although these prior measurements indicate that the magnetization of an array has quite different properties from the bulk constituent ferromagnetic materials, there has been little systematic attention paid to this fundamental collective property.

Here, we present a detailed experimental and simulation study of the magnetization of extended square ASI arrays, examining the temperature and field dependence of the magnetization of arrays with varied lattice spacing. We find behavior consistent with coherent magnetization reversal, closely following expectations of the Stoner-Wohlfarth model [31,32] for large lattice spacings, but with systematic deviations at small spacing. We compare our results with micromagnetic simulations, demonstrating that these deviations result from inter-island interactions that depend critically on individual island shape. Our results highlight a path to fine-tuning of the magnetic response of ASI and other nanomagnet arrays, with implications for both device applications [33,34] and novel collective magnetic states.

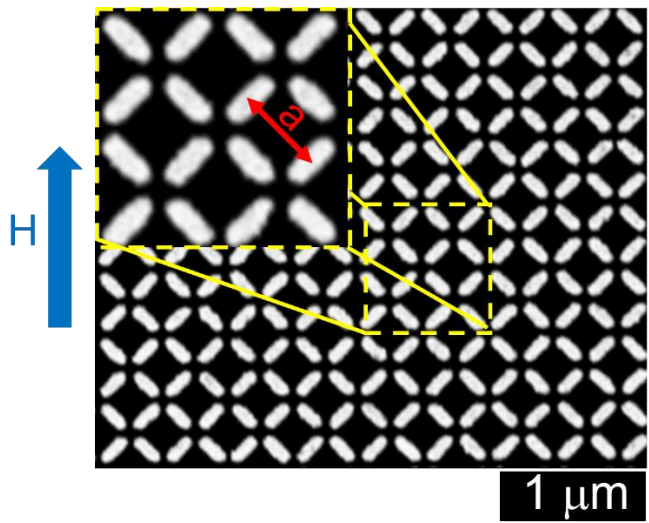

**Figure 1:** Scanning electron microscope image of a portion of a $t = 25$ nm sample (sample B), with lattice spacing $a = 320$ nm. The applied magnetic field ($H$) direction is shown.

Our permalloy (Ni$_{0.80}$Fe$_{0.20}$) square ASI arrays were patterned via electron-beam lithography (Figure 1) at thicknesses $t$ = 10, 15, and 25 nm, on Si/SiO$_x$ substrates, including two sets of $t$ = 25 nm samples, labeled A and B. Each array is composed of approximately 25 million islands, with lateral array sizes of 1 mm × 1 mm to 3.5 mm × 3.5 mm. Such large array sizes are required because of the extremely small moment of each island (~10$^{-13}$ emu). For each thickness, we studied arrays with varying lattice spacing ($a$ = 320, 380, 500, 1000 nm) as well as a separate set of samples (labelled C) with $t$ = 25 nm with $a$ = 280, 290, 300, 310, 320, 340, 360, 380 nm. These spacings correspond to distances from the end of the islands to the vertex center of 30 to 390 nm, for $a$ = 280 nm and $a$ = 1000 nm, respectively. The island size for all arrays was (220 ± 11) x (80 ± 8) nm. At these thicknesses, the Curie temperature ($T_C$) is at or near the bulk value (~850 K) [35], thus all measurements occur well below the superparamagnetic blocking temperature [36,37]. For comparison, we also measured a continuous permalloy film with $t$ = 25 nm. Samples were measured in a commercial SQUID magnetometer (MPMS3, Quantum Design), with the magnetic field ($H$) aligned 45° from the long-axis of the islands, so that all islands have the same orientation relative to $H$ (Figure 1) [17,22,24].

For temperature-dependent measurements, arrays were measured on warming after field-polarization at high temperature and cooling in zero field. We define the magnetization ($M$) as the measured magnetic moment normalized by the total number of islands. The saturation magnetization ($M_S$) is taken as the magnetization at +2 kOe, and the remanent magnetization ($M_R$) is the value at $H$ = 0 on decreasing the field from positive saturation. The coercive field ($H_C$) is taken as the $M$ = 0 crossing point averaged over positive and negative fields (or as the maximum slope in the hysteresis loop for the small number of samples with significant background contributions at $H_C \approx 0$, see Supplemental Material SM-5). We define $M_S$, $M_R$, and $H_C$ in Figure 2c. The values of magnetization are typically normalized to $M_S$ to account for lithographic defects, which introduce an uncertainty of ~10% in measured $M$ values (see Supplemental Material SM-1). Most data shown below are for $t = 25$ nm (Sample B), where such effects were minimized, and our results are consistent across all sample thicknesses.

In the inset of Figure 2a, we show $M_R(T)$ at various lattice spacings, where small vertical offsets among the curves are attributable to lithographic defects (see Supplemental Material SM-1). The temperature dependence of $M_R$ is considerable (~10%), even in this regime well below $T_C$. We note that the form of $M_R(T)$ is consistent among all lattice spacings, and with the continuous films, as shown in the normalized data in Figure 2a. This suggests that lateral dimensions of the islands and inter-island interactions do not substantially impact the thermally-excited spin dynamics that are responsible for the temperature dependence of $M_R$. While continuous permalloy films are very soft, with $H_C \approx 1$ Oe, the shape anisotropy of ASI islands leads to $H_C$ values of several hundred Oe [22,27,28], as shown in Figures 2b and 2c, where we illustrate the temperature evolution of $H_C$.

To further understand these temperature dependences, in Figure 2d we plot $\Delta M(T) = (M(25\text{ K}) - M(T))$ for both $M_R(T)$ and $M_S(T)$, as well as $\Delta H_C(T) = (H_C(25\text{ K}) - H_C(T))$. We show these data on a log-log scale, normalized to the low-temperature values, demonstrating a clear power-law dependence. Fits for $T > 100$ K (to avoid spurious effects from sidewall oxidation [38,39]) give $\Delta H_C(T) \propto T^{1.46\pm0.03}$, $\Delta M_S(T) \propto T^{1.56\pm0.04}$, and $\Delta M_R(T) \propto T^{2.08\pm0.02}$. Importantly, the proportionality of $\Delta H_C(T)$ and $\Delta M_S(T)$ is consistent with expectations from the Stoner-Wohlfarth model for coherent rotation of the island moments [11,40], in which

$$H_C \propto \frac{K}{M_S}, K \propto M_S^2$$

where $K$ is the (uniaxial) shape anisotropy constant. The first proportionality here arises from the Stoner-Wohlfarth model, while the second arises from shape anisotropy, assuming (as expected in permalloy) that it is dominant over magnetocrystalline anisotropy. Our measured exponents are consistent across all measured samples and lattice spacings (see Supplemental Material SM-5), and thus affirm previous evidence for coherent moment reversal in similar artificial spin ice systems [9,41].

We note that the common measured exponent for $\Delta H_C(T)$ and $\Delta M_S(T)$ is also consistent with expectations for a Bloch-like ($T^{3/2}$) dependence of the magnon-induced suppression of the magnetization in conventional ferromagnets [42]. The difference between the exponent values for $\Delta M_S(T)$ and $\Delta M_R(T)$ can be ascribed to the suppression of low-energy (long-wavelength) spin waves in an applied field, noting that finite dimensions are well known to affect spin waves and thus the temperature-dependent magnetization in nanoscale ferromagnets [43–50]. In addition, the difference between the exponent values for $M_R(T)$ and $M_S(T)$ is less dramatic in our continuous permalloy films, where we find $M_R \propto T^{1.88}$ and $M_S \propto T^{1.7}$. This again suggests that the field-suppressed spin waves in our ASI samples are those impacted by the lateral island dimensions. We note that the lowest-lying spin waves at the $\Gamma$ point in the first Brillouin zone of a square ASI are edge modes formed by oscillations of the magnetization localized near island edges [6]. These modes may lie below the uniform Kittel-like mode, and, in the presence of an external field, these modes contribute less to the thermal spin excitations that suppress the magnetization at finite temperatures.

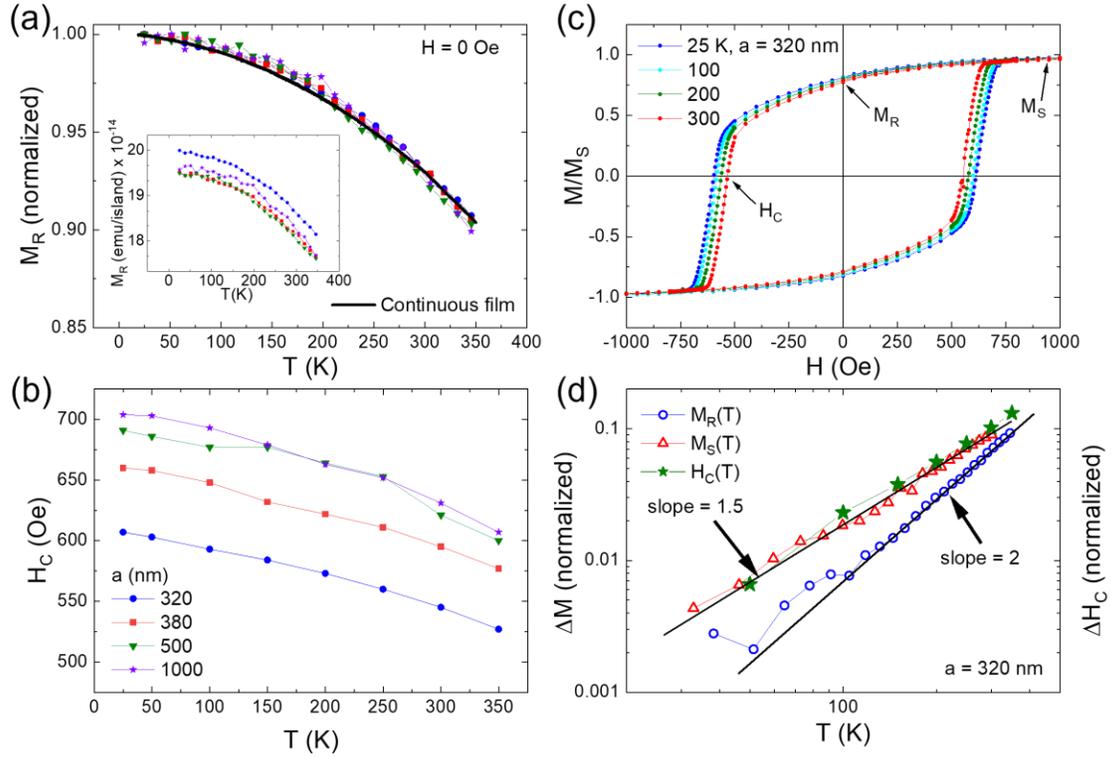

**Figure 2: (a)** Remanent magnetization, $M_R(T)$, normalized to the value at $T = 25$ K for various lattice spacings and a continuous film with $t = 25$ nm after each array was field-polarized at $T = 380$ K; Inset: the same $M_R(T)$ values without normalization. Legend presented in (b). **(b)** $H_C(T)$ for $a = 320, 380, 500$, and $1000$ nm. Small relative variation between the data for the two largest values of $a$ is well within the measurement uncertainty. **(c)** $M(H)$ for a variety of temperatures for $a = 320$ nm. **(d)** Log-log plots of $\Delta M(T) = (M_{T=25K} - M(T))$ for $\Delta M_S$ and $\Delta M_R$ and $\Delta H_C(T) = (H_{C,T=25K} - H_C(T))$ for $a = 320$ nm, normalized to the $T = 25$ K values of $M_S$, $M_R$, $H_C$ respectively. Solid lines show power laws, as described in the text. All data are for $t = 25$ nm, sample B.

In Figure 3a,b, we see that $H_C(a)$ is approximately constant for $a \geq 500$ nm and then decreases at smaller spacing, consistent with previous measurements [22]. Figure 3c shows that qualitatively similar lattice-spacing dependence is observed for all thicknesses, and that $H_C$ decreases with decreasing $t$, consistent with the expected smaller energy barrier to magnetization reversal. The three values of $t$ were chosen to validate the results for different moments and interaction strengths. The saturation of $H_C(a)$ at larger lattice spacing indicates that the behavior appropriately asymptotes to the case of non-interacting islands. The lattice spacing dependence at small spacing is consistent with expectations that interactions become most important in that limit. The slope of $H_C(a)$ increases substantially for the smallest lattice spacing, as expected due to the non-linear strength of dipolar interactions with island separation. The decrease in $H_C(a)$ continues down to our smallest measured lattice spacing of $a = 280$ nm, as shown in Supplemental Material SM-4 for sample C.

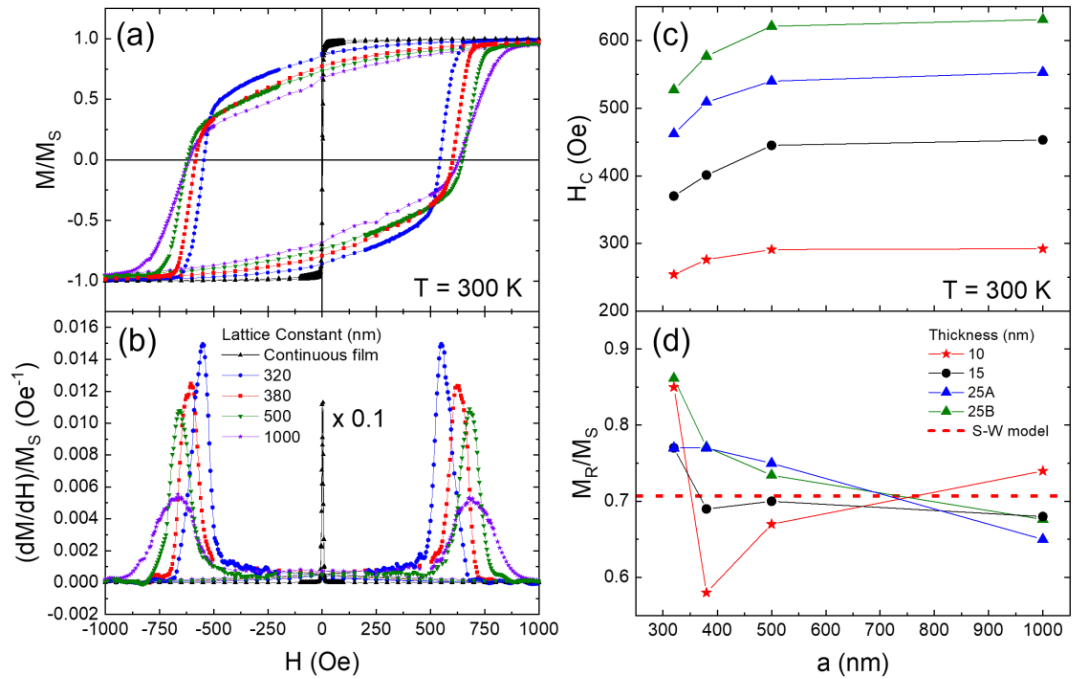

**Figure 3: (a)** $M(H)$ at 300 K for $t = 25$ nm sample B for a variety of $a$, including the 25 nm continuous film for comparison (legend shown in (b)). **(b)** Normalized derivative, $(dM/dH)/M_S$ of the data in (a), showing broadening of the loop with increasing $a$ (data smoothed using adjacent averaging over five neighboring points). **(c)** $H_C(a)$ at 300 K for multiple island thicknesses. **(d)** $M_R/M_S$ at 300 K as a function of $a$, with the Stoner-Wohlfarth value of 0.707 shown for comparison.

The shape of $M(H)$ in Figure 3a provides further evidence that the magnetization reversal in this system is qualitatively consistent with expectations of Stoner-Wohlfarth in the large-spacing limit [31,32]. The expected $M(H)$ for ideal Stoner-Wohlfarth behavior is shown in Figure 4a (details are given in Supplemental Material SM-8), where the qualitative similarity to the shape of the experimental hysteresis loops is notable, especially for $a$ = 1000 nm in Figure 3a. However, experimental loops become markedly more square by $a$ = 320 nm, suggesting that inter-island interactions induce deviations from coherent magnetization rotation, as might be expected [13,17]. These conclusions are reinforced by the behavior of $M_R/M_S$ shown in Figure 3d, where this "squareness ratio" is close to $\sim 1/\sqrt{2} = 0.707$, the Stoner-Wohlfarth value, but consistently higher for $a$ < 380 nm. Outliers in Figure 3d (e.g., for $t$ = 10 nm) are presumably associated with non-ideal $M(H)$ curves with steps at small $H$, in various samples (see supplemental material SM2 and SM5).

To better understand the magnetization reversal process, we performed athermal micromagnetic simulations (details given in Supplemental Material SM-7 and in reference [51]). Because there are many factors that cannot be reproduced precisely in micromagnetic simulations, including edge roughness and thermal fluctuations, we used the (temperature-dependent) $M_S$ as a fitting parameter to obtain $M(H)$ loops in good qualitative and semi-quantitative agreement with the experiment. For the results shown here, we used $M_S$ = 700 x $10^3$ A/m and assumed zero intrinsic magnetocrystalline anisotropy and a micromagnetic exchange parameter $A$ = 13 pJ/m [52] (the effects of changing $A$ are discussed in the Supplemental Material SM-7). Measurements at 300 K give $M_S$ = 733 x $10^3$ A/m for continuous 25-nm-thick films, and ~800 x $10^3$ A/m for bulk permalloy [47]. We used an island thickness $t = 25$ nm and varied the lattice spacing and shape. Simulated islands are oriented along the horizontal (*x*) and vertical *(y)* directions, with *H* applied 45° to the horizontal direction.

Previous micromagnetic simulations showed that the single island reversal process is strongly impacted by the shape of the island ends [10]. We therefore used two different island shapes in our simulations in Figure 4(d): S1 (rectangle with semicircular ends) and S2 (rectangle with elliptical ends), as detailed in Supplementary Material SM-6. Lithographic variations of our samples seen in SEM indicate variations of island shapes that do not perfectly map to either S1 or S2, however simulations for both shapes give values of $H_C$ within ~10% of the measured low-temperature value. As shown in Figure 4a, shape S2 generates good qualitative agreement with both experiments and the Stoner-Wohlfarth model with respect to the form of *M*(*H*). However, shape S1 has qualitatively different hysteresis loops (c.f., Figure 4b) that include a sharp change in slope, and small steps in the magnetization near $H_C$, confirming the sensitivity to island shape.

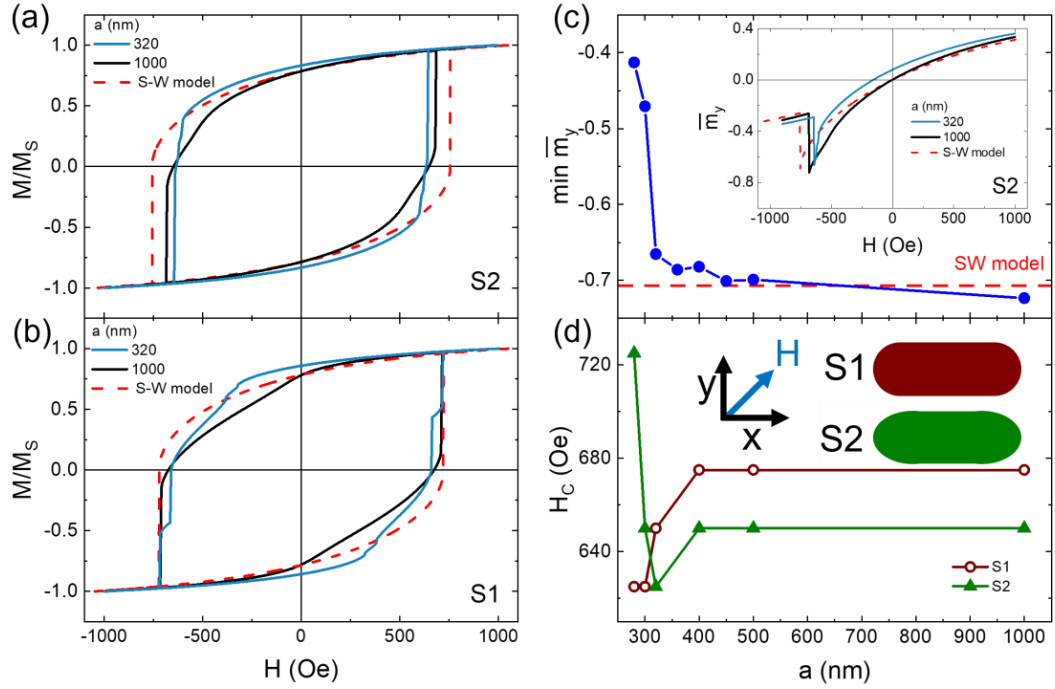

**Figure 4:** Simulated $M(H)$ (averaged over all islands) for $a$ = 320 and 1000 nm, and an ideal curve from the Stoner-Wohlfarth model for **(a)** shape S2 and **(b)** shape S1. **(c)** The minimum value of the y component of the average magnetization ($\bar{m}_y$) of horizontal islands as a function of magnetic field from +1000 Oe to -1000 Oe with the field at 45° to the *x* axis; dotted line shows the minimum value of $\bar{m}_y$ generated from the Stoner-Wolhfarth model. Inset: $\bar{m}_y(H)$ for the Stoner-Wohlfarth model and for a = 320 and 1000 nm for shape S2. **(d)** Simulated $H_C$ as a function of lattice spacing for shapes S1 and S2, and outlines of island shapes S1 and S2, showing the different end curvature.

To develop a more quantitative understanding, we plot $\bar{m}_y$, which is the vertical component of the magnetization, $m_y(H)$, for horizontal islands of shape S2 in the inset to Figure 4c, averaged over the full island. These data show a sharp minimum corresponding to magnetization reversal, with the depth of this minimum corresponding to how coherently the magnetization rotates. In the main panel of Figure 4c, we plot the depth of that minimum as a function of lattice spacing. We see that the minimum of $\bar{m}_y$ is near the expected value for the Stoner-Wohlfarth model at large spacings but rises significantly at smaller lattice spacings. This rise simply reflects that the magnetization has a more substantial twist from one edge of the island to the other during the reversal process (as seen in Figure 5), causing the loop to be more square in shape. This confirms that the island reversal is largely coherent in the absence of inter-island interactions.

While the hysteresis loops suggest that shape S2 is a more appropriate model to capture the behavior of our system, the effects of inter-island interactions indicate that a more nuanced understanding is required. [54] Figure 5 shows a mapping of the magnetization from simulations for both island shapes S1 and S2, at fields corresponding to $M_S$, $M_R$, $H_C$ for $a$ = 280, 320, and 1000 nm. The arrows show the local direction of the magnetization, and the color scale indicates the local value of $m_y$ for islands aligned along the $x$-axis. The effects of inter-island interactions are readily apparent from the spread in colors for the island near $H_C$, where a wider spread in colors implies less coherent rotation. Smaller lattice spacings create local fields, which impact rotation of the island magnetization near the island ends, thus making the reversal less coherent. We note that even though the magnetization is not fully coherent, the magnetization texture in islands of our studied size and shape is always smooth and not broken into distinct domains, which cannot be assumed for all cases [41].

Since inter-island interactions suppress coherence of magnetization reversal, we can now understand the reduced $H_C$ for smaller lattice spacings as a consequence of a smaller energy barrier for a less coherent reversal process in ASI. This is the case for both shapes S1 and S2, down to $a$ = 320 nm (Figure 4(d)). As $a$ is reduced below 320 nm, however, the simulated $M(H)$ of S1 and S2 arrays become markedly different, as shown in Figure 4d. $H_C$ increases dramatically with further reduced lattice spacing for shape S2, while $H_C$ decreases for S1 with decreasing lattice spacing. This difference between S1 and S2 can be understood in terms of the energy cost associated with the magnetization rotating near the curved edge of the islands. For the elliptical edges of S2, there is increased energy cost to rotate the magnetization, this effect is greatly enhanced by strong inter-island interactions, which couple to the more elliptical shape to create good flux closure horizontally and vertically at a vertex. Experimentally, microscopic island edge roughness presumably suppresses this energy cost, leading to behavior more like that of shape S1.

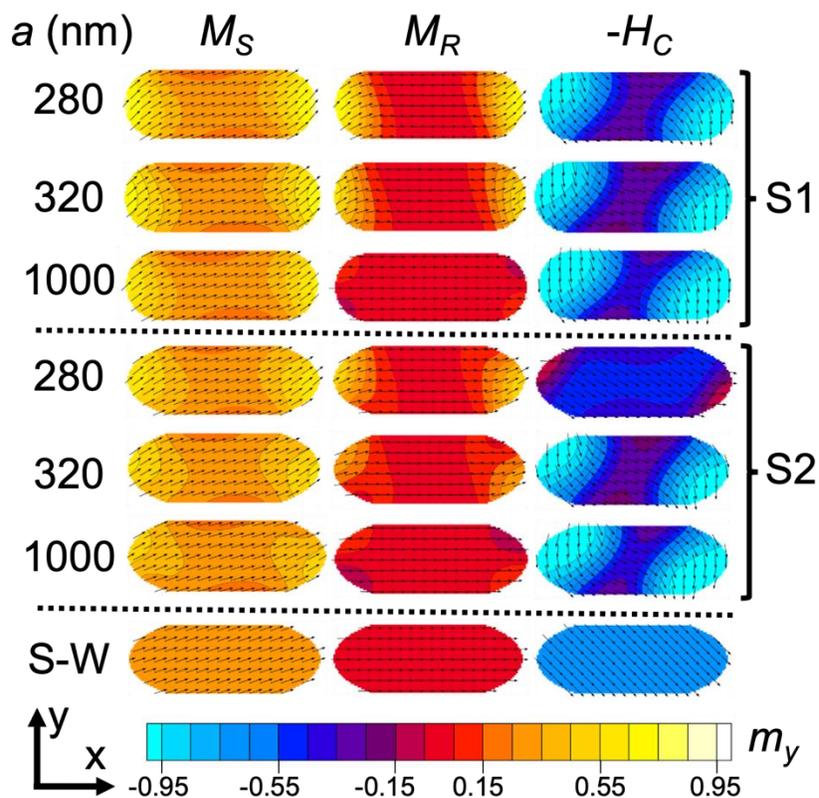

**Figure 5:** Simulated magnetization maps of individual islands for $a$ = 280, 320, and 1000 nm at $M_S$ ($H$ = 1000 Oe), $M_R$ ($H$ = 0), and $-H_C$ for island shapes **(top)** S1 ($H_C$ = -640 Oe, -660 Oe, and -710 Oe for 280, 320, and 1000 nm, respectively), **(middle)** S2 ($H_C$ = -735 Oe, -635 Oe, and -680 Oe for 280, 320, and 1000 nm, respectively), and **(bottom)** Ideal Stoner-Wohlfarth model mapped onto shape S2 ($H_C$ = -514.5 Oe). The values chosen for $H_C$ are just prior to the moment reversal. Arrows indicate the local direction of the magnetization, and color coding indicates its $y$-component $m_y/M_S$ with the field applied at 45° to the horizontal direction.

The substantial temperature and lattice spacing dependence in our measurements of $M$ and $H_C$ has direct implications for future studies in these systems, since fine-tuning of experimental protocols will need to take these effects into account. This is especially true for strongly interacting arrays where the simple expectation of coherent magnetization reversal following a Stoner-Wohlfarth model begins to break down. More generally, the results also have implications for the wide range of other interacting nanomagnet systems, including a range of superparamagnetic materials and patterned media that have important technological implications. Because the exact nature of the interaction-induced effects at low lattice spacing depend crucially on details of the shape of the islands, precise lithographic control could exploit this dependence to further control collective behavior. This creates the possibility for new degrees of freedom in tuning the behavior of these systems, both for applications and exploration of fundamental physics of collective phenomena.

## AUTHOR CONTRIBUTIONS AND FUNDING ACKNOWLEDGEMENT

Lithography and experimental measurements were conducted by NB, XZ, and PS at Yale University, funded by the US Department of Energy, Office of Basic Energy Sciences, Materials Sciences and Engineering Division under Grant No. DE-SC0020162. Sample growth was performed by JR and CL at the University of Minnesota and was supported by NSF through Grant No. DMR-2103711. Simulations were conducted by OH at Argonne National Laboratory with funding from the US Department of Energy, Office of Science, Basic Energy Sciences Division of Materials Sciences and Engineering. Data analysis and manuscript preparation was performed by NB, OH, CL, and PS. We gratefully acknowledge the computing resources provided on Blues, a high-performance computing cluster operated by the Laboratory Computing Resource Center at Argonne National Laboratory. We would also like to acknowledge Zhixin Zhang and Axel Hoffmann for useful discussions and verification. OH's present and permanent address is Seagate Technology, Bloomington, MN 55345.